\documentclass[twocolumn,showpacs,superscriptaddress,preprintnumbers,amsmath,amssymb]{revtex4}
\usepackage{graphicx,graphics,color,epsfig}
\usepackage{pdfpages}
\usepackage{bm}
\usepackage{amsmath}
\usepackage{amssymb}
\usepackage{epstopdf}
\usepackage[mathscr]{eucal}

\begin{document}
\bibliographystyle{prsty}
\title{Doping dependence of the electron-phonon and
electron-spin fluctuation interactions in the high-$T_{c}$ superconductor Bi$_{2}$Sr$_{2}$CaCu$_{2}$O$_{8+\delta}$}
\author{Elbert E. M. Chia}
\author{D. Springer}
\author{Saritha K. Nair}
\author{X. Q. Zou}
\author{S. A. Cheong}
\affiliation{Division of Physics and Applied Physics, School of
Physical and Mathematical Sciences, Nanyang Technological
University, Singapore 637371, Singapore}
\author{C. Panagopoulos}
\affiliation{Division of Physics and Applied Physics, School of
Physical and Mathematical Sciences, Nanyang Technological
University, Singapore 637371, Singapore}
\affiliation{Department of Physics, University of Crete and
FORTH, 71003 Heraklion, Greece}
\author{T. Tamegai}
\affiliation{Department of Applied Physics, The University of
Tokyo, Hongo, Bunkyo-ku, Tokyo 113-8656, Japan}
\author{H. Eisaki}
\affiliation{National Institute of Advanced Industrial Science
and Technology, Tsukuba 305-8568, Japan}
\author{S. Ishida}
\author{S. Uchida}
\affiliation{Department of Physics, The University of Tokyo,
Bunkyo-ku, Tokyo 113-0033, Japan}
\author{A. J. Taylor}
\author{Jian-Xin Zhu}
\affiliation{Los Alamos National Laboratory, Los Alamos, NM
87545, USA}
\date{\today}

\begin{abstract}
Using ultrafast optical techniques, we detect two types of bosons strongly coupled to electrons in the family of
Bi$_{2}$Sr$_{2}$CaCu$_{2}$O$_{8+\delta}$ from the
underdoped to overdoped regimes. The different doping dependences of the electron-boson coupling strengths enables us to identify them as phonons and spin fluctuations: electron-phonon coupling ($\lambda_{e-ph}$) peaks at optimal doping, and electron-spin fluctuation coupling ($\lambda_{e-sf}$) decreases monotonically with doping. This observation is consistent with two facts: (1) superconductivity is in close proximity with antiferromagnetism at low dopings, and (2) a pronounced lattice renormalization effect at larger dopings.
\end{abstract}

\maketitle

Despite many advances in understanding copper-oxide
high-transition-temperature ($T_{c}$) superconductors, there still exists no
universally accepted mechanism. Determining the nature of
interaction responsible for the Cooper-pair formation remains
one of the grand challenges in modern condensed matter physics.
The most probable candidates are lattice vibrations (phonons)
\cite{McQueeney99,Lanzara01}, spin fluctuation (SF) modes
\cite{Norman97}, and pairing without invoking glue
\cite{Anderson07}. For conventional superconductors, structure
in the electron tunneling $dI/dV$ characteristics established
unambiguously that the attractive pairing interaction was
mediated by phonons \cite{McMillan69}. For high-$T_{c}$
superconductors, structure in $dI/dV$ has also been found in
many tunneling measurements \cite{Kirtley07}. More recent
scanning tunneling microscopy (STM) experiments revealed an
oxygen lattice vibration mode whose energy is anticorrelated
with the local gap value on hole-doped Bi-2212 \cite{Lee06}
while a bosonic mode of electronic origin was found in the
electron-doped Pr$_{0.88}$LaCe$_{0.12}$CuO$_{4}$
\cite{Niestemski07}. Together with salient features observed in
angle-resolved photoemission spectroscopy (ARPES)
\cite{Lanzara01,Gweon04}, these new results raise the
fundamental question of whether the bosonic modes are a pairing
glue \cite{Balatsky06} or a signature of an inelastic tunneling
channel \cite{Pilgram06}.

The role of the electron-boson interaction in
high-$T_{c}$ superconductors has been studied by different
techniques. For example, inelastic neutron scattering tracks
the changes in boson energies or dispersions upon entering the
superconducting state. ARPES~\cite{ADamascelli03}
measure the effects of electron-boson interaction on
electronic self-energies, and planar junction experiments
determine the energy of the bosonic mode \cite{Zasadzinski01}.
STM measure the local density of states through the
local differential tunneling conductance, where the
characteristic boson mode energy is estimated from the dip
position \cite{Lee06}. However, it does not give the
electron-boson coupling strength directly because both the
coupling strength and mode energies are encoded in the electron
self-energy itself.

As for the electron-boson coupling strength in cuprates,
time-integrated optical measurements can be useful
\cite{Carbotte99,Hwang04a}, but it is difficult to elucidate
whether one or more bosons are involved.
Time-resolved pump-probe spectroscopy is a powerful technique
used to probe the relaxation dynamics of photoexcited
quasiparticles in correlated electron systems such as cuprate, pnictide and actinide superconductors, spin density wave materials, and Kondo systems \cite{Averitt02,Chia2006,Chia07,Chia2010,Burch08,Gadermaier10,Talbayev10}.
Its unique contribution lies in its ability to extract the
value of the electron-boson coupling strength ($\lambda$)
directly, via the electron-boson relaxation time, without the need to perform complicated inversion algorithms. This procedure has been experimentally verified on the conventional superconductors \cite{Brorson90}. In this Letter, we report measurements of time-resolved quasiparticle relaxation of high-quality single crystals of
underdoped (UD) to overdoped (OD) Bi$_{2}$Sr$_{2}$CaCu$_{2}$O$_{8+\delta}$ (Bi-2212, hole concentration $p$=0.10--0.22). Our data indicate the coupling of electrons to \textit{two} bosonic modes: the electron-phonon coupling constant ($\lambda_{e-ph}$) peaks at optimal doping, while the electron-SF coupling constant ($\lambda_{e-sf}$) decreases monotonically with doping.

The family of the bi-layer cuprate Bi-2212 has been the most
intensively studied class of high-$T_{c}$ superconductors in
recent years, due to their (a) extreme cleavability, (b)
containing only CuO$_{2}$ planes and not chains, and (c) the
possibility of growing samples with a larger range of
$T_{c}$'s. Single crystals of Bi-2212 were obtained from three
groups (two Tokyo and one Tsukuba) grown by the floating zone
method with doping control, yielding values of $T_{c}$
(determined by magnetization data) that depend on the hole
doping level ($p$) spanning from the UD ($p$=0.10,
$T_{c}$=65~K) to the OD ($p$=0.22, $T_{c}$=65~K)
regime. In the optimally-doped (OPT) sample, Ca has been doped
with Y to obtain the highest $T_{c}$ of 95~K. Underdoping was
achieved using excess Bi atoms substituted for the Sr sites as well
as reducing oxygen content, while the more OD samples have been
doped with Pb, to obtain lower values of $T_{c}$. The values of $p$
were obtained from the $T_{c}$ values using the parabolic law
\cite{Presland91} $T_{c}/T_{c}^{max}=1-82.6(p-0.16)^2$, where
$T_{c}^{max}$=95~K.

\begin{figure}
\includegraphics[width=8cm,clip]{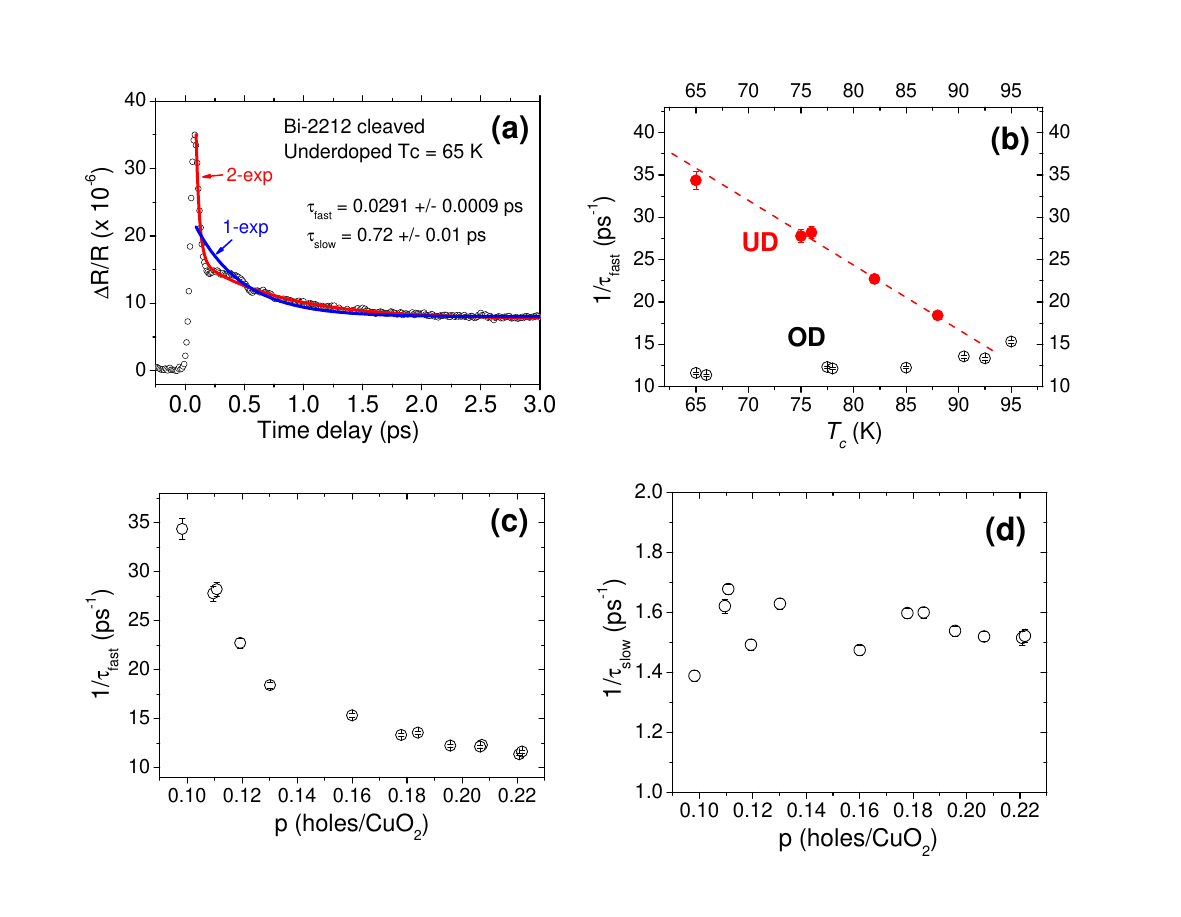}
\caption{(a) $\Delta R/R$ versus pump-probe time
delay, of an UD Bi-2212
sample ($T_{c}$=65~K). (o): Experimental data. Blue
line: one-exponential fit. Red line: two-exponential fit. (b) Fast relaxation rate $1/\tau_{fast}$ versus $T_{c}$. Note the difference in the strength and sign of the correlation in the OPT-OD and UD samples. Red dotted line is a guide to the eye. (c) $1/\tau_{fast}$ versus doping. In this representation $1/\tau_{fast}$ decreases monotonically with increasing doping, but with a faster decrease in the UD region. (d) $1/\tau_{slow}$ versus doping.}
\label{fig:Bi2212Fig1}
\end{figure}

To avoid any competing relaxation processes
from emergent low temperature states (e.g., superconducting,
pseudogap, antiferromagnetic, or stripe order), we
take all data at room temperature. Figure~\ref{fig:Bi2212Fig1}(a) shows the time dependence of the photoinduced change in reflection ($\Delta R/R$) of an UD Bi-2212 sample (see supplementary information \cite{Supp} for more discussion). The time evolution of the photoinduced reflectivity change $\Delta R/R$ first shows a rapid rise (of the order of the pump pulse duration) followed by a subsequent decay. As
shown in Fig.~\ref{fig:Bi2212Fig1}(a), the data can be fitted
better by two exponentials (red line) than a single exponential
(blue line). It indicates the quasiparticle relaxation has two
components --- the fast component $\tau_{fast}$ ($\sim$100~fs) and
the slow component $\tau_{slow}$ ($\sim$700~fs). We attribute
the fast relaxation process to the electrons first transferring
energy to a bosonic mode (e.g. phonons) which are more strongly
coupled at a characteristic time $\tau_{fast}$. These bosons
then continue to cool by energy dissipation via
anharmonic decay at a characteristic time $\tau_{slow}$.

Since the transfer of electron energy first occurs through
selected modes that are most strongly coupled to electrons, we
use $\tau_{fast}$ to be indicative of the electron-phonon
coupling strength. These strongly coupled phonon modes should
be the most relevant in discussing the possible phonon-mediated
superconductivity. Figure~\ref{fig:Bi2212Fig1}(b) shows the
$T_{c}$ dependence of the fast relaxation rate $1/\tau_{fast}$ --- there is a  \textit{weak and positive} correlation between $1/\tau_{fast}$ and $T_{c}$ for OPT to OD samples, while for the UD samples, the correlation is \textit{strong and negative},
in contrast to what is expected from the BCS theory, where $\lambda_{e-ph}$ (proportional to $1/\tau_{fast}$) correlates positively with $T_{c}$.  Figure~\ref{fig:Bi2212Fig1}(c) shows the
doping dependence of the fast relaxation rate $1/\tau_{fast}$ --- here $1/\tau_{fast}$ decreases monotonically with
increasing doping, with a faster decrease in the UD than the OD
region. These two panels suggest that, in the UD region, another important relaxation mechanism,
in addition to the electron-phonon interaction, contributes to the quasiparticle relaxation. For the slow relaxation rate $1/\tau_{slow}$, as shown in Fig.~\ref{fig:Bi2212Fig1}(d), it varies only slightly with doping, but a little more pronounced at low dopings.

\begin{figure}
\includegraphics[width=8cm,clip]{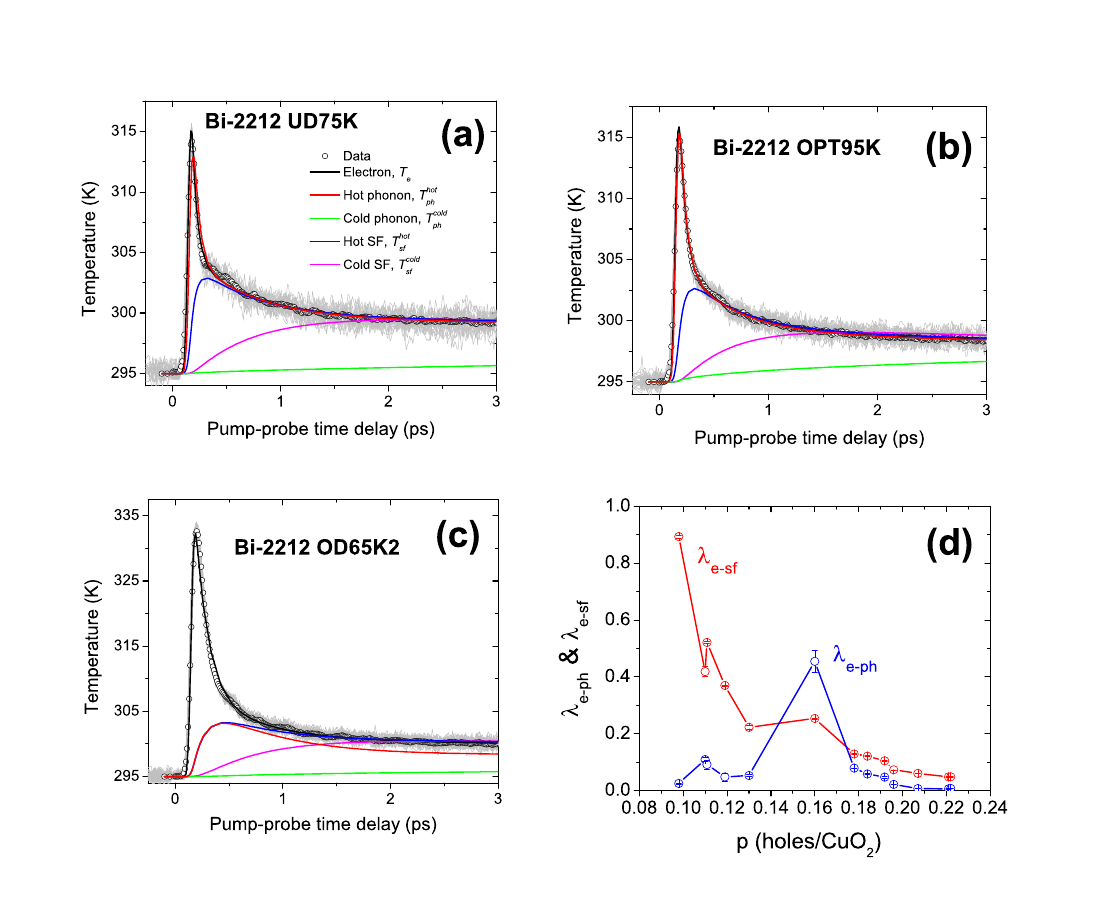}
\caption{Grey lines: Individual data sets. Open circles = averaged data, multiplied by a proportionality factor. The
proportionality factor [$(5.0\pm 0.1)
\times 10^{5}$~K] is a fitting parameter constrained
to be doping independent, since the same amount of
incident laser power is absorbed by all the samples. Solid
lines are time evolutions of the temperatures of the various
subsystems: electrons (black), hot phonons ($T^{hot}_{ph}$,
red), cold phonons ($T^{cold}_{ph}$, green), hot spin
fluctuations ($T^{hot}_{sf}$, blue), and cold spin fluctuations
($T^{cold}_{sf}$, magenta). All data, taken at 295~K, are shown here to 3~ps, but were taken and fitted up to 7~ps. (d) Doping dependence of
$\lambda_{e-ph}$ (blue) and $\lambda_{e-sf}$ (red). Fittings to the individual data sets yield a distribution of fitting parameters, and the standard deviations of these distributions give the error bars.}
\label{fig:Bi2212Fig3}
\end{figure}

We further elucidate the contribution of possible multiple
bosonic modes to our data and and extract their respective coupling strength
to electron based on an effective temperature model.
The existence of multiple relaxation rates in a single sample
suggests a conventional two-temperature mode is not sufficient.
We note that a three-temperature model (3TM) has been used
in the time-resolved ARPES analysis of an OPT Bi-2212
sample~\cite{Perfetti07}. In the 3TM,
photoexcited electrons first transfer their energy to the (``hot") phonons
that are more strongly coupled with them. These hot phonons
then lose their energy to the cold phonons through anharmonic
coupling. This model has been successfully used to fit the data for one single sample OPT Bi-2212 in Ref.~\onlinecite{Perfetti07}.
However, to interpret the relaxation phenomena across the whole range of dopings as presented here, we need to go beyond the 3TM. Especially,
the presence of strong SFs in the UD regime leads
us to consider a five-temperature model (5TM). In this new model,
there exists a second cooling channel for the hot electrons,
that is, via coupling to hot SFs, which
subsequently also cool via scattering with cold SFs. The time evolution of the temperatures of the various sub-systems: electronic ($T_{e}$), hot/cold phonon ($T_{ph}^{hot}$/$T_{ph}^{hot}$), hot/cold SFs ($T_{sf}^{hot}$/$T_{sf}^{hot}$) satisfy the rate equations:
\begin{equation}
\frac{dT_e}{d\tau} = \frac{P}{C_e} - C_1 \frac{n^{ph}_{e} - n_{ph}^{hot}}{T_e} - \frac{3\lambda_{sf}(\Omega^{hot}_{sf})^3}
{\hbar\pi k_{B}^{2}} \frac{n^{sf}_{e} - n_{sf}^{hot}}{T_e}
\label{eqn:5TM1}
\end{equation}
\begin{equation}
\frac{dT^{hot}_{ph}}{d\tau} = \frac{C_e}{C^{hot}_{ph}} C_1 \frac{n^{ph}_{e}
- n_{ph}^{hot}}{T_e} - \frac{T^{hot}_{ph} - T^{cold}_{ph}}{\tau_{ph}}
\label{eqn:5TM2}
\end{equation}
\begin{equation}
\frac{dT^{hot}_{sf}}{d\tau} = \frac{C_e}{C^{hot}_{sf}}
\frac{3\lambda_{sf}(\Omega^{hot}_{sf})^3}{\hbar\pi k_{B}^{2}}
\frac{n^{sf}_{e} - n_{sf}^{hot}}{T_e} - \frac{T^{hot}_{sf} -
T^{cold}_{sf}}{\tau_{sf}}
\label{eqn:5TM3}
\end{equation}
\begin{equation}
\frac{dT^{cold}_{ph}}{d\tau} =
\frac{C^{hot}_{ph}}{C^{cold}_{ph}}\frac{T^{hot}_{ph} - T^{cold}_{ph}}{\tau_{ph}}
 \label{eqn:5TM4}
\end{equation}
\begin{equation}
\frac{dT^{cold}_{sf}}{d\tau} =
\frac{C^{hot}_{sf}}{C^{cold}_{sf}}\frac{T^{hot}_{sf} - T^{cold}_{sf}}{\tau_{sf}}
 \label{eqn:5TM5}.
 \end{equation}
The system is excited by a Gaussian pulse $P$ with FWHM 45 fs and energy density of 0.166 J/cm$^{3}$. The specific heat of electrons, hot/cold phonons, and hot/cold SFs are
$C_{e} = \gamma T_{e}, C_{ph}^{hot}=N_{ph}f_{ph}\Omega_{ph}^{hot} \frac{\partial n_{ph}^{hot}}{T_{ph}^{hot}}, C_{ph}^{cold}=N_{ph}(1-f_{ph})\Omega_{ph}^{cold} \frac{\partial n_{ph}^{cold}}{T_{ph}^{cold}}, C_{sf}^{hot}=N_{sf}f_{sf}\Omega_{sf}^{hot} \frac{\partial n_{sf}^{hot}}{T_{sf}^{hot}}$, and $C_{sf}^{cold}=N_{sf}(1-f_{sf})\Omega_{sf}^{cold} \frac{\partial n_{sf}^{cold}}{T_{sf}^{cold}}$, respectively. The parameter $f_{ph}$ ($f_{sf}$) denotes the fraction of total phonon (SF) modes that are more strongly coupled to the electrons, and $N_{ph}$ ($N_{sf}$) denotes the number of phonon (SF) modes in the irradiated volume. The distribution functions are $n_{ph}^{hot(cold)} = [e^{\Omega_{ph}^{hot(cold)}/k_{B}T_{ph}^{hot(cold)}}-1]^{-1}$,  and
$n_{sf}^{hot(cold)} = [e^{\Omega_{sf}^{hot(cold)}/k_{B}T_{sf}^{hot(cold)}}-1]^{-1}$. In Eq.~(\ref{eqn:5TM1}), $n_{e}^{ph(sf)} = [e^{\Omega_{ph(sf)}^{hot}/k_{B}T_{e}}-1]^{-1}$ are not distribution functions, but are results of performing delta-function energy integrals \cite{Allen87}.

The energy of the hot phonon mode $\Omega^{hot}_{ph}$=40~meV corresponds to the out-of-plane out-of-phase oxygen buckling $B_{1g}$ phonon. Though cuprate
samples like Bi-2212 are inhomogeneous both in energy gap and characteristic boson frequency, the spatial average of mode frequency is doping independent \cite{Lee06}. Therefore, we assume $\Omega^{hot}_{ph}$ is constant throughout the entire doping regime. Neutron scattering data \cite{Renker89} reveal
an accoustic phonon mode at 20~meV. Since recent STM data \cite{Lee06} did not observe any coupling between electrons and this particular accoustic mode, we choose it to be the energy of our cold phonon bath ($\Omega^{cold}_{ph}$). The hot ($\Omega^{hot}_{sf}$) and cold ($\Omega^{cold}_{sf}$) SF energies are taken from optical spectroscopy \cite{Hwang07} and are doping dependent: both $\Omega^{hot}_{sf}$ and $\Omega^{cold}_{sf}$ scale with the centroid position of the broad bosonic background, with the constraint that $\Omega^{hot}_{sf}$=41~meV at OPT. Our model also takes into account the fact that (a) the samples change from a good
metal at higher doping to a bad metal at lower dopings (via factor $C_{1}$) \cite{Supp,Kabanov08}, and (b) the penetration depth at 800~nm pulse changes with doping \cite{Hwang07}. Two other parameters do not change with doping: (a) the Sommerfeld coefficient $\gamma$ \cite{Junod90,Loram04} and (b) fraction of incident laser power absorbed by the sample \cite{Supp}. The two most important parameters to be extracted from this model are the coupling constants $\lambda_{e-ph}$ and $\lambda_{e-sf}$.

Figure ~\ref{fig:Bi2212Fig3}(a)--(c) shows the 5TM fits of three representative samples. The time evolution of the various sub-systems (electron, hot/cold phonons/SFs) depends on the relative strengths of the coupling constants \cite{Supp}. The optimal hot phonon and SFs were obtained using a parameter space study \cite{Supp}. The change in peak shapes is well described by our model, and we obtained very good fits for the entire doping range. Figure~\ref{fig:Bi2212Fig3}(d) shows the doping dependence of the respective coupling constants. In the UD region, $\lambda_{e-ph}$ initially increases with doping between $p$=0.098 and 0.11, then starts to decrease between 0.11 and 0.13 doping, before increasing again to a peak at OPT ($p$=0.16), and thereafter decreasing with increasing doping in the OD region. On the other hand, $\lambda_{e-sf}$ decreases with increasing doping, with a stronger decrease in the UD than the OD regions. We also performed fits using other possible hot phonon and SF modes: (a) half-breathing in-plane copper-oxygen bond stretching phonon as the hot phonon mode, with $\Omega^{hot}_{ph}$=70~meV, (b) magnetic resonance mode as the hot SF mode, with doping-dependent energy values obtained from neutron scattering data \cite{Fauque07}, and/or the peaks in the bosonic spectrum in optical spectroscopy data \cite{Hwang07,Supp}. Only the magnitudes of $\lambda_{e-ph}$ and $\lambda_{e-sf}$ differ; the overall trends with doping do not change.

The monotonic decrease of $\lambda_{e-sf}$ with doping is consistent with the weakening of SFs with increasing doping, resulting in a weaker coupling between electrons and SFs. Note the strong decrease of $\lambda_{e-sf}$ occurs up to at $p$=0.13; thereafter the decrease becomes more gradual. This change in behavior at $p$=0.13 could be related to the formation of stripe ordering \cite{Kivelson03} at $p$=1/8. The non-zero value of $\lambda_{e-sf}$, even in our most OD sample ($p$=0.22), suggests that SFs are present even in the OD regime, in agreement with studies by inelastic neutron scattering \cite{Lipscombe07}.

For $\lambda_{e-ph}$, its initial increase between 0.098 and 0.11 doping is qualitatively consistent with BCS theory, where an increased electron-phonon coupling gives rise to an increased $T_{c}$ and vice versa. Its subsequent decrease between 0.11 and 0.13 ($\sim$1/8) doping, increase away beyond $p$$\sim$1/8, may be due to the presence of stripe order at this doping level \cite{Watanabe00}.  The concurrent decrease of $\lambda_{e-sf}$ in this same ``stripe'' region implies a suppression in the superconductivity, consistent with a plateau in the superfluid density near this 1/8 doping \cite{Anukool09}. The doping
dependence of $\lambda_{e-ph}$ in the OPT-OD region --- maximum at OPT, and decrease with overdoping, is also in qualitative agreement with BCS theory, and strong lattice renormalization effects \cite{Lanzara01,Cuk04}.

Independent confirmation of the electrons coupling to multiple bosonic modes come from frequency-resolved and time-resolved pump-probe measurements of OPT Bi-2212 \cite{Conte12}. The resulting spectral function necessitates the electrons to be coupled directly to hot phonons, cold lattice, as well as a bosonic mode of electronic origin, which the authors suggested could be SFs or current loops. Moreover, ARPES data on heavily UD La$_{2-x}$Sr$_{x}$CuO$_{4}$ showed fine structure in the electron self-energy, demonstrating the involvement of multiple boson modes in the coupling with electrons \cite{Zhou05}. The contribution of our work, besides working on a different class of cuprate superconductors, is that we have obtained the \textit{doping dependence} of the
different electron-boson coupling constants from the UD to the OD regimes.

It is important to realize that our 5TM requires \textit{both} the electron-phonon and electron-SF interactions to be responsible for the fast ($\sim$100~fs) relaxation. Important confirmation of this comes from the temperature-dependent pump-probe data of our most OD sample ($T_{c}$$\sim$65 K) \cite{Nair10}. Besides observing the opening up of a pseudogap at $T^{\ast}$$\sim$100~K, the fast relaxation time $\tau_{fast}$ increases with decreasing temperature, before peaking at $T^{\ast}$ and decreasing to $\sim$100~fs at 30~K. This change in behavior of $\tau_{fast}$ at $T^{\ast}$ is intriguing --- it suggests that the electron-SF coupling, in addition to electron-phonon coupling, is involved in the initial fast relaxation of the hot electrons. The peak at $T^{\ast}$, and its subsequent decrease below $T^{\ast}$, is then due to an increased scattering rate between electrons and SFs as the sample enters the pseudogap phase. This scenario is further confirmed by the temperature-dependence of $\tau_{fast}$ above $T^{\ast}$ --- a fit to $1/T^{n}$ yields $n$=1.3, which disagrees with the behavior predicted for the electron-phonon relaxation time for good ($n=2$) and poor ($n=3$) metals \cite{Kabanov08}.

In conclusion, we have performed ultrafast optical techniques on a wide range of doping levels in Bi-2212. The relaxation analysis suggests the existence of {\em two} types of bosonic modes strongly coupled to electrons. The different doping dependences of the electron-boson coupling strengths enables us to identify them as phonons and spin fluctuations: electron-phonon coupling ($\lambda_{e-ph}$) peaks at optimal doping, and electron-spin fluctuation coupling ($\lambda_{e-sf}$) decreases monotonically with doping. The observation should shed new insight into the mechanism of high-$T_c$ superconductivity in cuprates.

E.E.M.C. acknowledges useful discussions with D. Mihailovic. This work was carried out under the auspices of the NNSA of the U.S. DOE at LANL under
Contract No. DE-AC52-06NA25396, MEXT-CT-2006-039047 and EURYI, the Singapore MOE AcRF Tier 1 (RG41/07) and Tier 2 (ARC23/08), and the National Research Foundation of Singapore (NRF-CRP4-2008-04).

\bibliography{Bi2212,Bi2212OD70K}

\begin{widetext}


\section*{Supplementary material}

**Supplementary Information for**
**Doping dependence of the electron-phonon and electron-spin fluctuation interactions in the high-$T_c$ superconductor $Bi_2Sr_2CaCu_2O_{8+\delta}$**

In this EPAPS auxiliary online document, we provide (A) details of experimental pump-probe setup, (B) general comments on the temperature models, (C) a comparison between a five- and a three-temperature model, (D) doping-dependent and independent parameters in the 5TM, (E) curve-fitting procedure, (F) time evolution of the electronic, phononic, and spin fluctuation subsystems based on the fits, (G) details about the error analysis and (H) doping dependences of the electron-phonon and electron-spin fluctuation coupling constants using three other scenarios of the phonon and spin fluctuation mode energies.

### A. Femtosecond pump-probe setup

In our experiments an 80-MHz repetition rate Ti:sapphire laser produces 45-femtosecond pulses at approximately 800 nm (~1.5 eV) as the source of both pump and probe optical pulses. The pump and probe pulses were cross-polarized, with a pump spot diameter of ~60 μm and probe spot diameter of ~30 μm. The reflected probe beam was focused onto an avalanche photodiode detector. The pump beam was modulated at 1 MHz with an acoustic-optical modulator to minimize noise. The experiments were performed with an average pump power of 3 mW, giving a pump fluence of ~0.1 μJ/cm$^2$ and a photoexcited quasiparticle density of 0.02/unit cell, showing that the system is in the weak perturbation limit. The probe intensity was ~10 times lower. Resolution is at least 1 part in $10^6$. The fitted values of τ have a typical error of ±1%. All data were taken at room temperature. For all the dopings in this paper, the same amount of the incident laser power is absorbed by the sample, i.e. ~90%.

### B. Temperature Model

On the timescale of several tens of femtoseconds (fs), it is *not inappropriate* to describe the various sub-systems with their corresponding temperatures. In the theoretical paper by Allen [1], he suggested a way to calculate the electron-phonon coupling constant ($\lambda_{e-ph}$) from the electron-phonon relaxation time ($\tau_{e-ph}$) in conventional superconductors. Brorson [2] followed up on his work by fitting pump-probe data of many conventional superconductors to the two-temperature model, and obtained values of the $\lambda_{e-ph}$ that agree very well with the literature. This shows that the effective temperature description of electrons is reasonable even in time scales of ~100 fs. In the cuprate superconductors, the on-site Hubbard interaction is $U = 8t \approx 1.04$ eV, where $t$ is the hopping amplitude [Refer to Ref. [3] to get $t$ = 130.5 meV]. This corresponds to an electron-electron thermalization time $\tau_{e-e}$ of ~4 fs (using $\tau \sim h/U$, $h$ being the Planck's constant). Therefore, owing to this large $U$, the electron-electron interaction is much stronger than that in conventional superconductors. This $\tau_{e-e}$ is much smaller than electron-boson long-tail relaxation time. As such, it will take a much shorter time for the photoexcited electrons to equilibrate to attain a Fermi-Dirac distribution. We maintain therefore, that it is reasonable to talk of an effective electron temperature in the cuprates at the sub-100 fs timescale. Perfetti *et al.* also used the concept of electronic temperature in the interpretation of his time-resolved ARPES data [4]. Finally, the extracted electron-phonon relaxation timescale (sub-100 fs) itself makes the effective temperature internally consistent.

### C. Comparison: Five-Temperature (5TM) vs. Three-Temperature Model (3TM)

Following Perfetti *et al.* we started out fitting our reflectivity data to a three-temperature model (3TM), where photo-excited electrons relax back into the ground state via electron-phonon interaction. In this section we explain how the 3TM failed to fit our data.

When using the 3TM to describe multi-level electron relaxation processes, we assume that the measured reflectivity of the sample is proportional to the electronic temperature. This proportionality constant should be the same for all samples, and thus we do not treat it as a fitting parameter. Instead, we first tune this proportionality constant by hand for a sample near optimal doping, to pick a value that would enable us to get good fits. As it turned out, we can pick one value for this proportionality constant ($p_1$), so that the tail of the electronic temperature-delay can be fitted out to 7 ps, at the expense of a good fit around the peak electronic temperature (see Figure 1). For another value of the proportionality constant ($p_2$), we can obtain a good fit around the peak electronic temperature, but will no longer be able to fit the tail well out to 7 ps (see Figure 2).

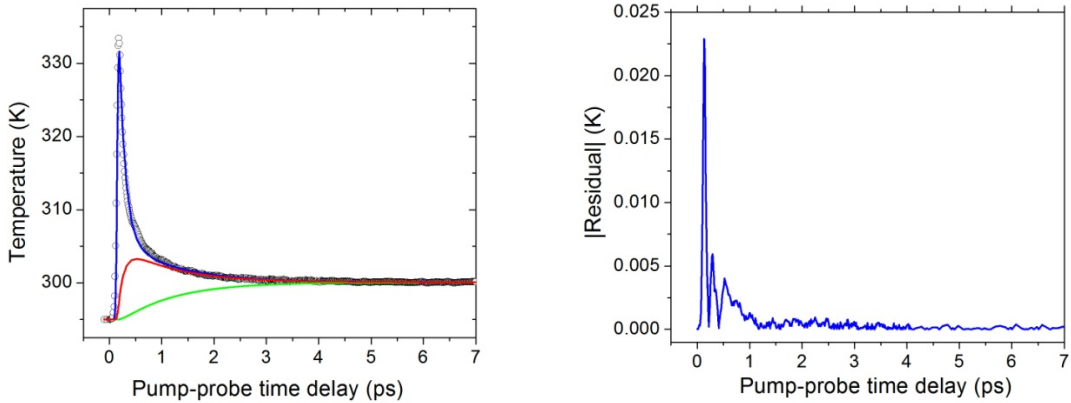

**Figure 1. (Nearly Optimal Doping).** The fit shown in the left figure underestimates the peak. At the tail the three temperatures become equal i.e. the electronic system is in equilibrium with the cooling phonon bath (blue dotted line: experimental data multiplied by proportionality constant, black line: fitted electronic temperature, red line: hot phonon temperature, green line: cold phonon temperature). The right figure shows the absolute of the residuals at each point of the curve. The residuals become smaller at later times.

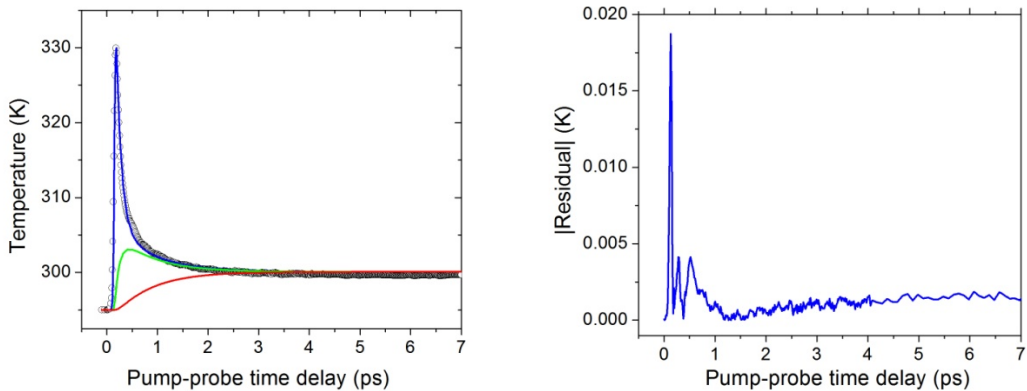

**Figure 2. (Nearly Optimal Doping).** The fit is in very good agreement around the peak (left). Equilibrium between the electronic and bosonic systems is reached around 2 ps. All temperatures remain constant from then on and thus cooling at later times is not captured. In comparison to Figure 1 we find that residuals (right) are now smaller around the peak but increase at later times.

When we fit the ultrafast data of the other samples to the 3TM, using the proportionality constant $p_1$, we find the fits are poor both around the peak and the tail (see for example, the fit to an overdoped sample in Figure 3). The same is true if we use the proportionality constant $p_2$ (see for example, the fit to the same overdoped sample in Figure 4).

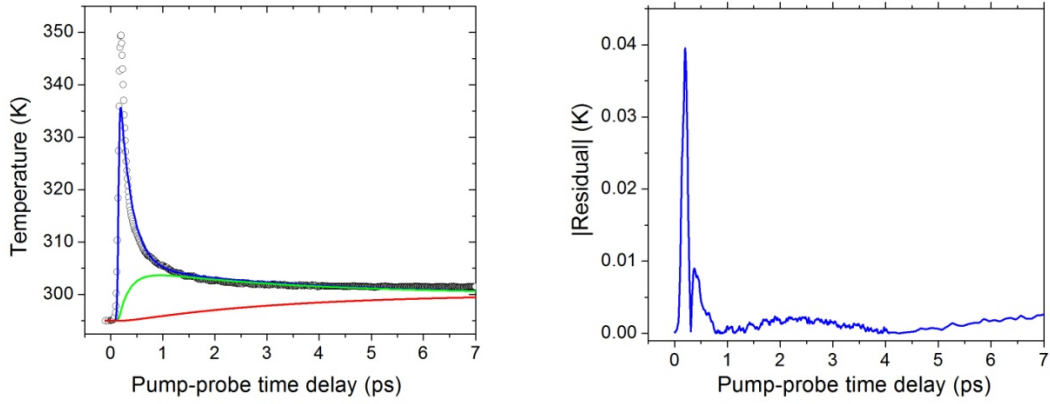

**Figure 3. (Overdoped Sample).** If we use the same ratio between reflectivity and temperature as in Figure 1 for overdoped samples we find that the peak is not estimated properly. From the residuals (right) and from the actual fit (left) it can be seen that the tail is not properly captured as well.

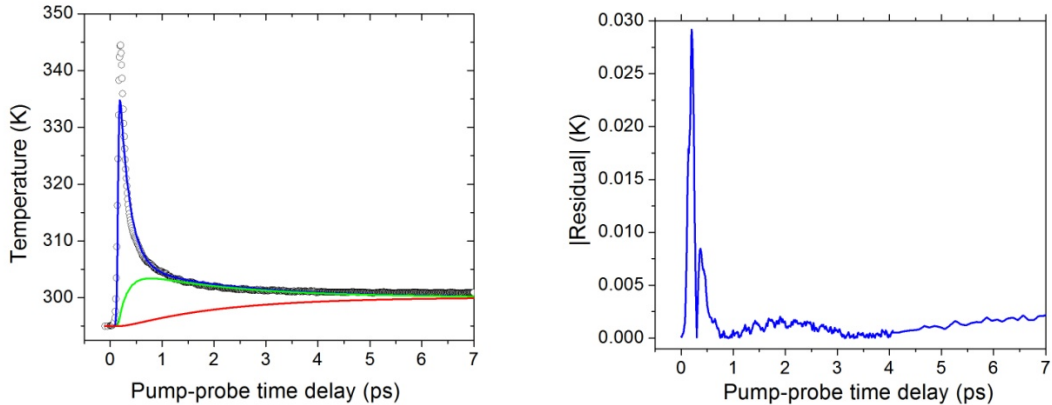

**Figure 4. (Overdoped Sample).** If we use the same ratio between reflectivity and temperature as in Figure 2 we get an improved tail section but the peak still is strongly underestimated.

In fact, no choice of proportionality constant will produce an acceptable compromise across the series of samples between fitting the peak electronic temperature, and also fitting the tail out to 7 ps. This suggests that we need cooling channels beyond the one within the 3TM to fit the peak and the tail well across all dopings. We found that the 5TM, which contains two cooling channels, is the simplest extension of the 3TM that admits good fits to the peak and the tail across all dopings using a common proportionality constant.

**D.     Doping-dependent and doping-independent parameters in the 5TM equations**

**(1)     Change in metallicity as a function of doping**

In Kabanov's paper [5], the spectral function $\alpha^2 F(\omega) \propto \omega^2$ in good metals, while in poor metals $\alpha^2 F(\omega) \propto \omega$. In the time evolution of the electronic temperature $T_e$, this effect gives rise to the prefactor $\left( \dfrac{3\lambda_{ph}(\Omega_{ph}^{hot})^3}{\hbar \pi k_B^2} \right)$ in good metals, and $\left( \dfrac{\pi^3 \lambda_{ph} k_B T_e^3}{2\hbar} \right)$ in poor metals. This moderation of the electron-phonon interaction, by the change in metallicity with doping, has been accounted for in Equations (1) and (2), by assuming a "two-fluid model", where $C_1$ is the linear combination of these two prefactors:

$$C_1 = (1-\alpha)\left( \frac{\pi^3 \lambda_{ph} k_B T_e^3}{2\hbar} \right) + \alpha\left( \frac{3\lambda_{ph}(\Omega_{ph}^{hot})^3}{\hbar \pi k_B^2} \right) \tag{17}$$

with the "good-metal" fraction $\alpha$ given by

$$\alpha(p) = a + bp. \tag{18}$$

We assume that $\alpha$ varies linearly with doping level $p$. This linearity is justified by the approximate linear dependence, with doping, of the residual resistivity of Bi-2212 [6] --- see Figures 5 and 6. This "good-bad-metal" picture, though phenomenological, is consistent with our picture of Bi-2212 becoming more metallic with increasing doping. In the underdoped region, the hole carriers are so few that the transport is difficult. At high temperatures, the hole carriers are scattered off the random distributed spin moments so strongly that the resistivity is much larger, i.e. "bad" metal. In the overdoped region, the effective bandwidth is increased and the spin fluctuations are suppressed significantly. As such, the resistivity is decreased, i.e. "good" metal.

The parameters $a$ and $b$ are determined by insisting that

$$\alpha(0.05) = 0 \quad \text{and} \quad \alpha(0.27) = 1, \tag{19}$$

implying that the material is a bad metal at $p = 0.05$ and a good metal at $p = 0.27$. These two end-points correspond to the dopings on the superconducting dome when the superconducting transition temperature $T_c = 0$. Figure 1 of Ref. [7] shows that, even at 300 K, the temperature where we took our data, these two end-points are still reasonable.

**(2)     Doping-dependent penetration depth**

Optical reflectance data of Bi-2212, at 300 K, showed a doping dependence of the penetration depth at 12500 cm$^{-1}$, corresponding to the wavelength of the pump and probe pulses (800 nm) [6].

**(3)     Doping-independent electronic specific heat**

In Equation (6), the Sommerfeld coefficient $\gamma = 8$ mJ/mol.K$^2$ from Junod [8]. There is no doping dependence of $\gamma$ for Bi-2212 at room temperature, as seen from specific heat data of Loram *et al.* [9].

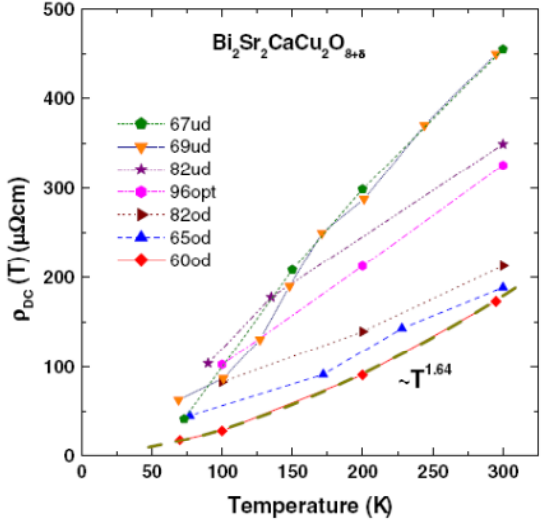

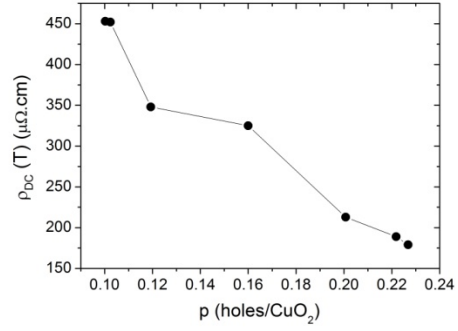

**Figure 5.** The dc resistivity, $\rho_{dc}(T)$, extracted from low-frequency extrapolations of normal state optical conductivities. The dashed curve through the data points for the most overdoped sample is a fitted curve, $\rho_{dc}(T) = 0.0155\,T^{1.64}$. Taken from Hwang [6]. Permission granted from IOP Publishing Limited.

**Figure 6.** Resistivity (at 300 K) versus doping, obtained from Figure 5. The trend is roughly linear, which justifies the use of Eqn. (18).

## E.    Curve-fitting procedure

For the results presented in this section the bosonic mode energies are chosen as described in our main article: Hot phonon energy [10] = 40 meV, cold phonon energy [11] = 20 meV while both the hot and cold spin fluctuation energies scale with the centroid position of the broad bosonic background [6], with the constraint that the energies are 41 meV at optimal doping. We start the fitting by assigning initial values to the parameters ($\lambda_{e\text{-sf}}$, $\lambda_{e\text{-ph}}$, $\tau_{sf}$, $\tau_{ph}$, $f_{sf}$, $f_{ph}$). Based on these values the system of five coupled ordinary differential equations is solved and the five effective temperatures ($T_e$, $T_{ph}^{hot}$, $T_{ph}^{cold}$, $T_{sf}^{hot}$, $T_{sf}^{cold}$) are obtained as functions of time $t$. We then calculate the sum of square errors between the electronic temperature ($T_e$) and the experimental data over a range of 7 ps. We then search through the parameter space to find the optimum set of parameters (minimum sum of square-errors) using the MATLAB built-in function *fminsearch*.

Our fitting results based on six fitting parameters revealed that eventually not all six parameters are doping dependent. To investigate this in more detail we begin with an unrestricted fit to identify those parameters with unambiguous doping dependence. We found these to be the coupling functions $\lambda_{e\text{-sf}}$ and $\lambda_{e\text{-ph}}$, and that the quality of the fits depends most sensitively on their values.

In the unrestricted fits we further found that the parameters $f_{sf}$, $f_{ph}$ do not show strong doping dependences. If we replace them with constants, and perform restricted fits to the remaining four parameters ($\lambda_{e\text{-sf}}$, $\lambda_{e\text{-ph}}$, $\tau_{sf}$, $\tau_{ph}$), we find the fitting errors of each individual sample change the least compared to the unrestricted fit across all samples. To determine the optimal values for the fractions, we perform a parameter space study. We calculate the fitting error of the 5TM with four fitting variables ($\lambda_{e\text{-sf}}$, $\lambda_{e\text{-ph}}$, $\tau_{sf}$, $\tau_{ph}$) at all possible fraction combinations ($f_{sf}$, $f_{ph} = 0.01 - 0.96$ at

5

0.05 intervals). This gives us one error surface for each of the differently doped samples. We then average over all of them to obtain the final error surface shown in Figure 7.

We found two minima in the averaged error surface that lie at almost symmetric locations in the parameter space. This is expected since the two bosonic channels are of similar form. The channels are distinguishable by their mode energies. At the fraction pairs that produce minima in the averaged error surface, all the individual error surfaces have a pronounced minimum as well. Thus the choice of fractions at the minimum of the averaged error surface is also a very good choice for each sample.

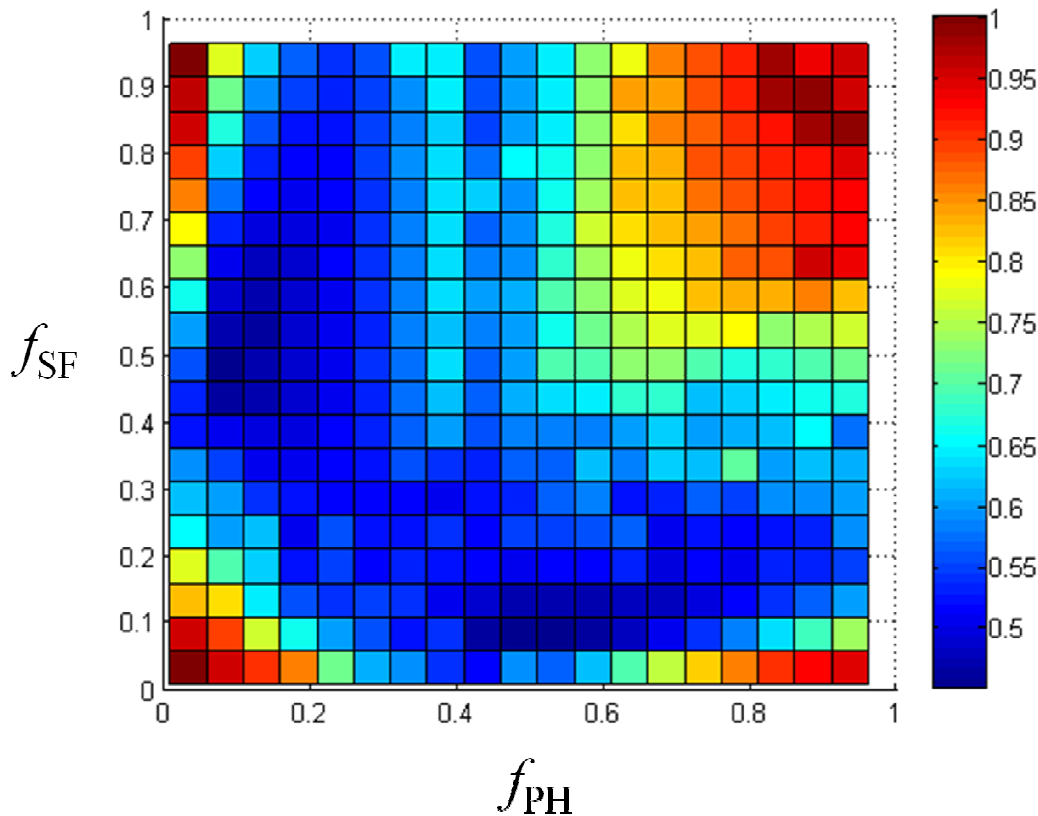

**Figure 7: Intensity Plot of averaged error surface.** The averaged error surface features two minima at symmetric locations along the $f_{sf}$ and $f_{ph}$ axis respectively. Such symmetry is expected because of the similarity of the bosonic cooling channels. A dark-blue coloring denotes fraction combinations that yield very good fitting results. The dark-red regions indicate large fitting errors. The locations of the minima are $(f_{ph}, f_{sf}) = (0.06, 0.46)$ and $(0.46, 0.06)$.

From the shape of the error surface two more conclusions can be drawn. First, the fact that the two minima are located away from the diagonal suggests that the two bosonic channels play essentially different roles. If a single bosonic cooling channel, like in a three temperature model, were sufficient, the error surface would feature minima along the diagonal ($f_{sf} = f_{ph}$) only. We found that the channel with a large fraction acts fast and equilibrates with the electronic temperature within the first picoseconds ($0 - 3$ ps). The channel with a small fraction of hot modes controls the cooling process at later times ($3 - 7$ ps). Only a combination of both cooling

channels allows good fitting results of up to 7 ps. Secondly, if the fractions were redundant degrees of freedom the error surface would not show sharp localized minima. Instead, a flat featureless surface would be expected. Our study thus confirms that both cooling channels should consist of one hot bath and one cold bath.

For the fraction pair $(f_{ph}, f_{sf}) = (0.06, 0.46)$, the doping dependence of the coupling constants is shown in Figure 2(d) of the main article. $\lambda_{e\text{-}ph}$ exhibits a peak at OPT, while $\lambda_{e\text{-}sf}$ decreases with increasing doping. For the other fraction pair $(f_{ph}, f_{sf}) = (0.46, 0.06)$, not only are the doping dependences of $\lambda_{e\text{-}ph}$ and $\lambda_{e\text{-}sf}$ reversed, the fits are also not as good as the first case. Since, with increasing doping, one is "farther" from the antiferromagnetic region, we should expect a decreasing $\lambda_{e\text{-}sf}$. We thus take the doping dependence of the coupling constants based on the first fraction pair, that is, $(f_{ph}, f_{sf}) = (0.06, 0.46)$.

## F.  Time evolution of electronic, phononic and spin fluctuation subsystems

Figure 8 reproduces Figures 2(a)-(c) of the manuscript. After the initial photoexcitation, the electrons heat up, rapidly transferring their energy to other subsystems. Due to the small number of hot phonons ($f_{ph} = 0.06$), equilibration with electrons ($T_{hot}^{ph} = T_e$) is reached within the first picosecond. The value of $\lambda_{e\text{-}ph}$ determines how fast the hot phonons respond to the electrons, and characterizes the sharpness of $T_{hot}^{ph}$-peak — from UD to OPT $\lambda_{e\text{-}ph}$ increases, resulting in a sharper hot-phonon peak; in the OD samples the hot phonons react slower due to a smaller $\lambda_{e\text{-}ph}$. Hot spin-fluctuations ($T_{hot}^{sf}$), on the other hand, equilibrate with electrons within the first 2 ps, due to their larger population ($f_{ph} = 0.46$) compared to hot phonons. The subsequent cooling of the electronic ($T_e$) peak is thus mainly determined by the coupling between electrons and hot spin-fluctuations. With increasing doping, $\lambda_{e\text{-}sf}$ decreases, which is reflected in a less-sharp $T_{hot}^{sf}$-peak. Between 1.5 to 3 ps, cold spin-fluctuations ($T_{cold}^{sf}$) equilibrate with electrons. Beyond 3~ps (hot and cold) spin-fluctuations are not capable to decrease $T_e$ any further — further electronic onwards cooling is due to cold phonons only. At times >7 ps all five temperatures eventually equilibrate and the tail flattens out. The small shoulder at ~0.5 ps is an extrinsic effect present in all UD samples, but does not affect the quality of the fits.

## G.  Error Analysis

To estimate the fitting error of our parameters we repeat the same fitting process for a large number of experimental data sets. While the actual parameters are obtained by fitting the average of the experimental data, the respective error bars are estimated by fitting all individual data sets separately. We then took standard deviations of the distributions of the various fitting parameters to obtain the error bars.

Figure 8 shows the original experimental data as grey shaded area behind the averaged curve used to obtain the actual fitting parameters. When we fit the wide range of noisy data we observe that only a narrow range of fitting parameters is necessary to obtain good fits. This suggests that our model parameters react sensitively to changes in the shape of the curve which is an indicator for a tightly constrained model.

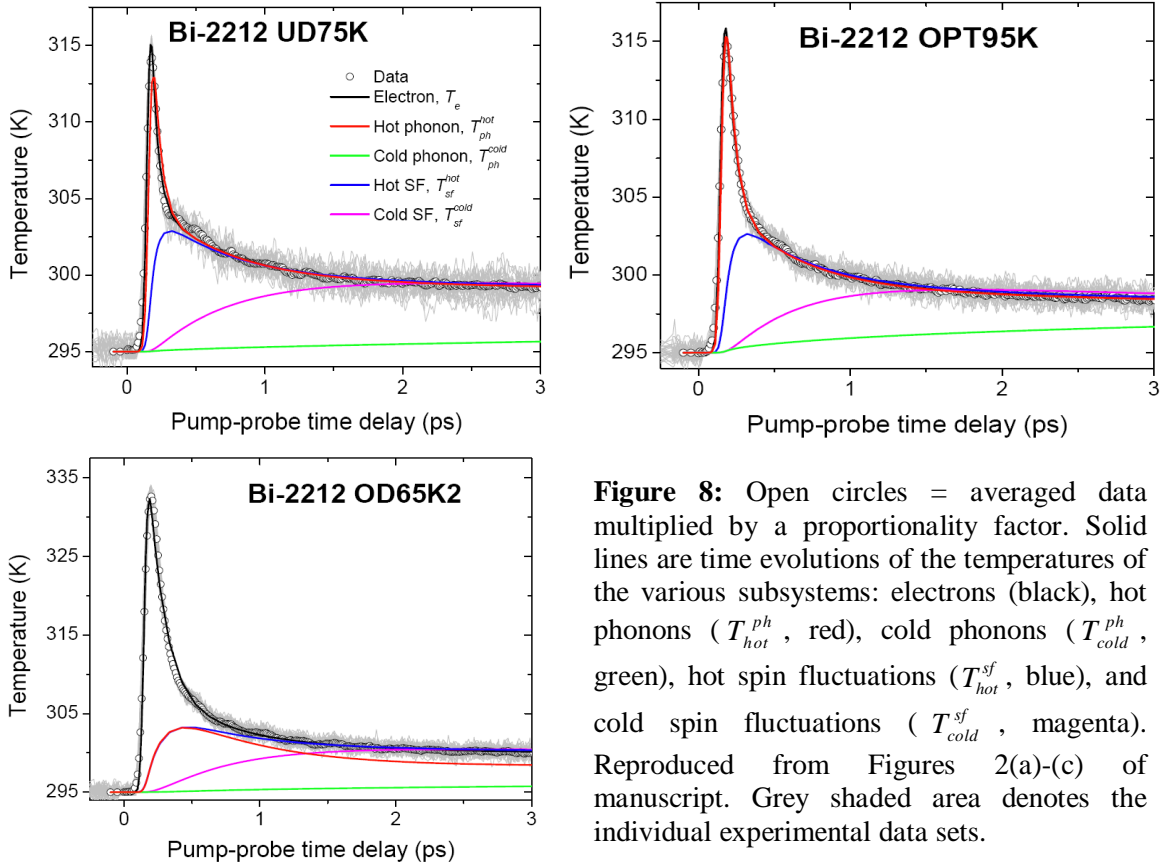

**Figure 8:** Open circles = averaged data multiplied by a proportionality factor. Solid lines are time evolutions of the temperatures of the various subsystems: electrons (black), hot phonons ($T_{hot}^{ph}$, red), cold phonons ($T_{cold}^{ph}$, green), hot spin fluctuations ($T_{hot}^{sf}$, blue), and cold spin fluctuations ($T_{cold}^{sf}$, magenta). Reproduced from Figures 2(a)-(c) of manuscript. Grey shaded area denotes the individual experimental data sets.

### H. Other choices of hot phonon and hot spin fluctuation energies

Besides the scenario presented in the main text, we show here the doping dependences of $\lambda_{\text{e-ph}}$ and $\lambda_{\text{e-sf}}$ for three other combinations of hot phonon and hot spin fluctuation energies (Figure 9). The fraction pairs used for these scenarios were obtained from error surface studies as well.

(a)   Hot phonon energy = 40 meV [10]. Hot spin fluctuation energies taken from the energies of the magnetic resonance mode from neutron scatterig data [12], and the positions of the peak in the bosonic spectrum from optical spectroscopy data [6].

(b)   Hot phonon energy = 70 meV [10]. Hot spin fluctuation energies taken from the energies of the magnetic resonance mode from neutron scatterig data [12], and the positions of the peak in the bosonic spectrum from optical spectroscopy data [6].

(c)   Hot phonon energy = 70 meV [10]. Hot spin fluctuation energies taken from the centroid of the broad background of the bosonic spectrum in optical spectroscopy data [6].

As shown in Figure 9, the doping dependence of $\lambda_{\text{e-ph}}$ and $\lambda_{\text{e-sf}}$ is quite robust against whichever scenario as listed above is used.

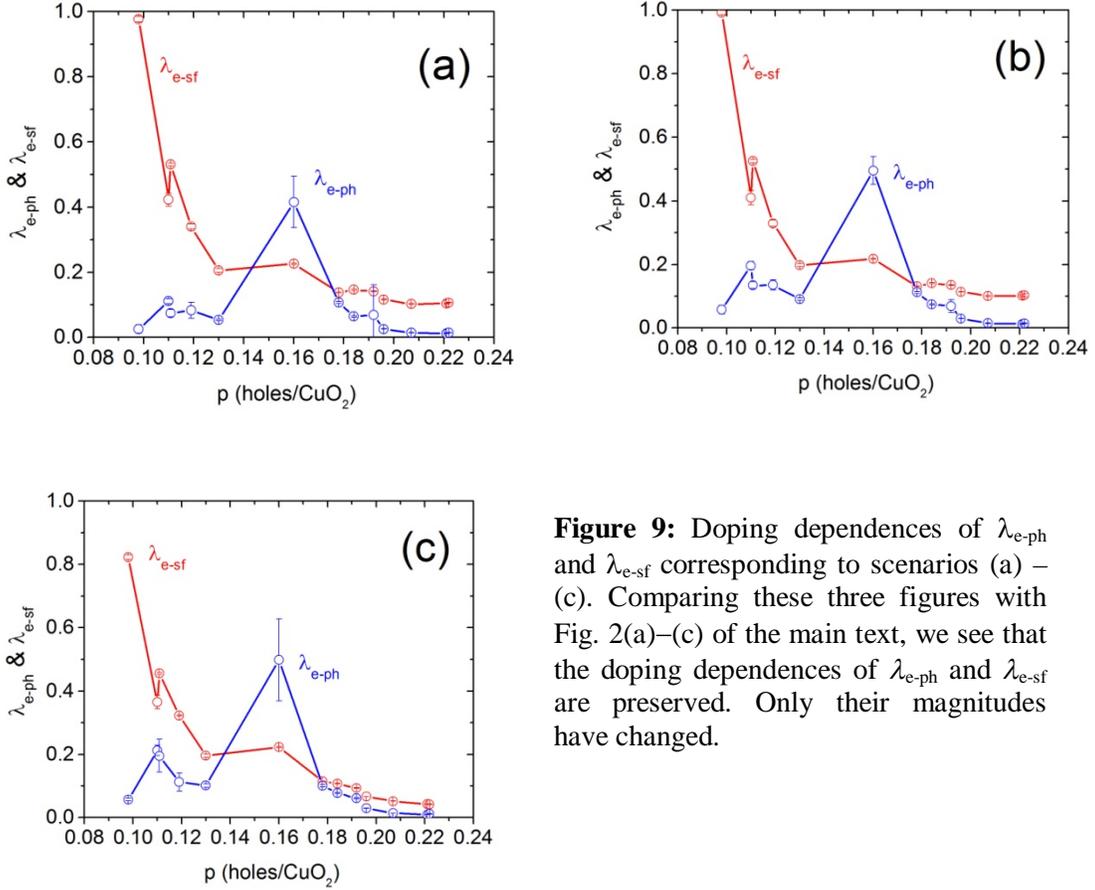

**Figure 9:** Doping dependences of $\lambda_{\text{e-ph}}$ and $\lambda_{\text{e-sf}}$ corresponding to scenarios (a) – (c). Comparing these three figures with Fig. 2(a)–(c) of the main text, we see that the doping dependences of $\lambda_{\text{e-ph}}$ and $\lambda_{\text{e-sf}}$ are preserved. Only their magnitudes have changed.

The doping dependence of $\lambda_{\text{e-ph}}$ and $\lambda_{\text{e-sf}}$, obtained from our fits looks physically reasonable — $\lambda_{\text{e-ph}}$ is roughly BCS-like, peaking at optimally doping, while $\lambda_{\text{e-sf}}$ gets stronger while approaching the antiferromagnetic insulator region. We consider this an indirect proof of our model. If the five-temperature model were not valid, then there is no reason, *a priori*, that the doping dependence of $\lambda_{\text{e-ph}}$ and $\lambda_{\text{e-sf}}$ would look the way they appear. In particular, in the underdoped region, most of the electrons are localized as local moments, which are responsible dominantly for the spin fluctuations, while the doped holes are effectively low-energy charge-carrying quasiparticles. The differentiation of the spin and charge degrees of freedom leads to the effective quasiparticle-spin fluctuation coupling, in addition to the electron-phonon coupling. More importantly, we were encouraged to suggest the five-temperature model by the fact that this new model interprets the experimental data nicely throughout the wide doping region considered in the present work.


%\includepdf[pages=1-?]{SIarXiv.pdf}
\end{widetext}

\end{document}